\documentclass[twocolumn,secnumarabic,amssymb, nobibnotes, aps,showpacs,prd]{revtex4}
\usepackage{epsfig}
\usepackage{amsmath,dsfont}
\usepackage{overpic}

\begin{document}
\title{\bf The {\boldmath $p\bar p$} mass threshold structure in 
{\boldmath $\psi(3686)$} radiative decay revisited}
\vskip 1.5true cm
\author{J.~Haidenbauer$^{1}$}
\author{Ulf-G. Mei{\ss}ner$^{1,2}$}
\affiliation{$^{1}$Institute for Advanced Simulation, Institut f\"ur
  Kernphysik, and J\"ulich Center for Hadron Physics,\\
Forschungszentrum J\"ulich, D-52425 J\"ulich, Germany\\
$^{2}$Helmholtz Institut f\"ur Strahlen- und Kernphysik and
Bethe Center for Theoretical Physics,\\ Universit\"at Bonn, D-53115 Bonn, Germany}

\begin{abstract}
The near-threshold behavior of the $p{\bar p}$ invariant mass spectrum
from the $\psi(3686){\to}\gamma p{\bar p}$ decay reported recently 
by the BESIII Collaboration is analyzed.
The enhancement in the $p{\bar p}$ invariant mass spectrum near threshold 
is nicely reproduced by the $p{\bar p}$ final-state interaction based on
the isospin averaged $^{1}S_0$ partial-wave amplitude as predicted by the 
J\"ulich nucleon--antinucleon model. 
Contributions from the $f_2(1910)$ or $f_2(1950)$ mesons, as promoted in
earlier works, are not needed. 
\end{abstract}


\maketitle

Recently, the BESIII Collaboration presented data with improved
statistics on the $p{\bar p}$ invariant mass spectrum for the 
reaction $J/\psi {\to}\gamma p{\bar p}$, 
but also a first high-statistics measurement of the 
$\psi(3686) {\to}\gamma p{\bar p}$ decay~\cite{BES12}.
The new $J/\psi {\to}\gamma p{\bar p}$ measurement confirmed
the spectacular near-threshold enhancement in the $p{\bar p}$ invariant 
mass, found in an earlier experiment by the BES Collaboration \cite{BES03}, 
which has been seen as evidence for a 
$p{\bar p}$ bound state or baryonium~\cite{Datta,Ding1,Suzuki,Dedonder},
for exotic glueball states~\cite{Chua,Rosner},
but also simply as manifestation of the final-state interaction (FSI) 
between the outgoing proton and antiproton
\cite{Sibirtsev05,Sibirtsev06,Kerbikov,Bugg,Zou,Loiseau,Entem}. 
A significant near-threshold enhancement of the $p{\bar p}$ invariant
mass was seen also in the $\psi(3686)$ decay, although less 
pronounced than in the $J/\psi$ case. 

The BESIII Collaboration themselves interpreted their $J/\psi$ decay data in 
terms of the $p{\bar p}$ FSI proposed by us \cite{Sibirtsev05}, but folded 
with a Breit-Wigner type resonance at around $1835$ MeV presuming 
that the structure at the $p{\bar p}$ threshold might be related to the 
$X(1835)$ resonance that had been observed 
in the reaction $J/\psi{\to}\gamma\pi^+\pi^-\eta'$ \cite{BES05,BES11},
see also the comments in Ref.~\cite{Sibirtsev06}.
This object is called $X(p{\bar p})$ in Ref.~\cite{BES11}.
For the description of the $\psi(3686)$ decay the same $X(p{\bar p})$
amplitude is used but sizeable additional contributions from the $f_2(1910)$ 
resonance had to be invoked. Contributions of a tensor meson, but in this case
of the $f_2(1950)$, were also advocated in the work of the CLEO collaboration
\cite{CLEO10}, which had published data on the reaction
$\psi(3686) {\to}\gamma p{\bar p}$ decay prior to BESIII, though with
lower statistics.
In both cases an isoscalar meson with a larger mass
($f_0(2100)$ and $f_2(2150)$, respectively) has been added to explain the
$p{\bar p}$ spectrum at higher invariant masses. 
 
In this report we take a closer look at those
$\psi(3686){\to}\gamma p{\bar p}$ data from the BESIII Collaboration.
Specifically, we provide an alternative interpretation of the
near-threshold enhancement solely in terms of the $p{\bar p}$ FSI,
i.e. without resorting to any resonance contributions like the $f_2(1910)$ 
or $f_2(1950)$ (or the $X(1835)$),
based on the very same $N\bar N$ interaction used by us previously in the 
explanation of the enhancement in the $J/\psi$ decay \cite{Sibirtsev05}. 
 
Conservation laws for parity, charge-conjugation and total
angular momentum severely restrict the partial waves in the $p{\bar p}$ 
system \cite{Sibirtsev05} for such decay processes. Specifically,
the partial-wave analysis for the $J/\psi$ decay performed in 
\cite{BES12} suggests that the near-threshold enhancement is dominantly
in the $J^{PC} = 0^{-+}$ state, which means that the $p{\bar p}$ 
system should be in the $^{1}S_0$ partial wave 
(we use here the standard nomenclature $^{(2S+1)}L_J$
where $S$ is the total spin and $L$ the orbital angular momentum). 
However, since the decay of the $J/\psi$ and $\psi(3686)$
to the $\gamma p{\bar p}$ system involves electromagnetic processes,
isospin is not conserved so that, in principle, any combination of the 
isospin $I=0$ and $I=1$ components is allowed. 
Indeed, while the $p{\bar p}$ invariant mass for $J/\psi$ decay
can be understood in terms of the FSI generated by the isospin $I=1$ 
component of the $N{\bar N}$ amplitude in the $^{1}S_0$ state alone 
-- at least in our work \cite{Sibirtsev05} -- the $I=1$ and $I=0$ channels 
can occur with different weights in case of the $\psi(3686)$ decay. 
 
The $\psi(3686) {\to}\gamma p{\bar p}$ decay rate is given 
by \cite{Sibirtsev05} 
\begin{eqnarray}
d\Gamma = \frac{|A|^2}{2^9 \pi^5 m_{\psi}^2}\,
\lambda^{1/2}(m_{\psi}^2,M^2,m_{\gamma}^2) \nonumber \\
\times\lambda^{1/2}(M^2,m_p^2,m_p^2)\, dM d\Omega_p\,  d\Omega_\gamma,
\label{spectr}
\end{eqnarray}
where the K\"all\'en function $\lambda$ is defined by
$\lambda (x,y,z)={((x-y-z)^2-4yz})/{4x}\,$,
$M \equiv M_{p\bar p}$  is the invariant mass of the $p{\bar p}$
system, $\Omega_p$ is the proton angle in that system,
while $\Omega_\gamma$ is the $\gamma$ angle in
the $\psi(3686)$ rest frame. 
After averaging over the spin states and
integrating over the angles, the differential decay rate is
\begin{eqnarray}
\frac{d\Gamma}{dM}=\frac{(m_{\psi}^2-M^2)
\sqrt{M^2-4m_p^2}}{2^7 \pi^3 m_{\psi}^3}\,\, |A|^2 \ .
\label{trans}
\end{eqnarray}
The quantity $A$ in Eqs.~(\ref{spectr}) and (\ref{trans}) stands for
the total $\psi(3686) {\to}\gamma p{\bar p}$ reaction amplitude and
is dimensionless. 

We assume again the validity of the Watson-Migdal \cite{Watson,Migdal}
approach for the
treatment of the FSI effect. It suggests that the reaction amplitude
for a production and/or decay reaction that is of short-ranged
nature can be factorized in terms of an elementary (basically
constant) production amplitude $A_0$ and the $p\bar p$ scattering
amplitude $T$ of the particles in the final state so that
\begin{eqnarray}
A (M_{p \bar p}) \approx A_0 \cdot N \cdot T(M_{p \bar p})~,
\label{fsi}
\end{eqnarray}
where $N$ is an arbitrary normalization factor, see e.g. Ref.~\cite{Sibirtsev05} 
for further details. As in our investigation 
of the $J/\psi$ decay we employ the amplitudes predicted by 
the $N\bar N$ model A(OBE) published in Refs.~\cite{Hippchen,Mull}, 
and we assume that the FSI effects in the $\psi(3686)$ decay are
likewise dominated by the $^{1}S_0$ partial wave.

\begin{figure}[t]
\psfig{file=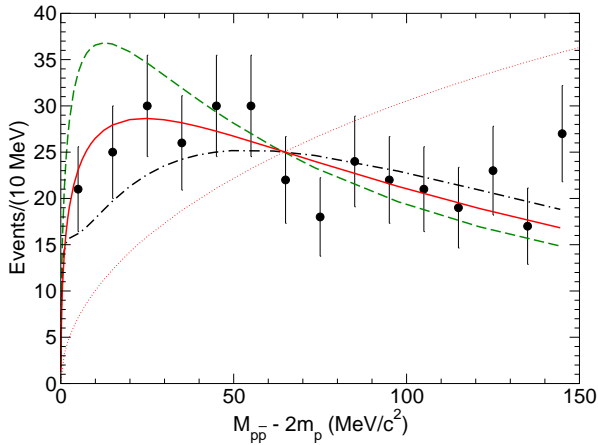,height=8.5cm,angle=-90}
\caption{
The $p\bar p$ mass spectrum from the decay $\psi(3686){\to}\gamma p{\bar p}$.
The circles show experimental results of the BES Collaboration
\cite{BES12}. The solid line is a calculation using the $p\bar p$ scattering 
amplitude squared ($|(T_{I=1}+T_{I=0})/2|^2$) predicted by 
the $N\bar N$ model A(OBE) \cite{Hippchen,Mull} for the $^{1}S_0$ partial wave, 
appropriately normalized to the data, cf. Eq.~(\ref{fsi}). The dashed 
(dash-dotted) curve are corresponding results using the $I=1$ ($I=0$)
amplitudes alone while 
the dotted line is the spectrum obtained from Eq. (\ref{trans})
by assuming a constant reaction amplitude $A$. Those curves were normalized 
so that they all coincide at $M_{p\bar p}-2m_p\approx$ 60 MeV.
}
\label{fig1}
\vspace*{-5mm}
\end{figure}

\begin{figure}[t]
\psfig{file=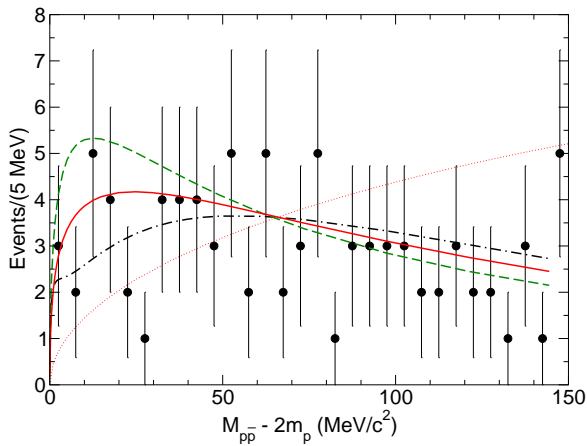,height=8.5cm,angle=-90}
\caption{
The $p\bar p$ mass spectrum from the decay $\psi(3686){\to}\gamma p{\bar p}$.
The circles show experimental results of the CLEO Collaboration
\cite{CLEO10}. For notation of curves, see Fig.~\ref{fig1}.
}
\label{fig2}
\vspace*{-5mm}
\end{figure}

Our results are presented in Fig.~\ref{fig1} together with the
data from Ref.~\cite{BES12}. As expected from the curves presented 
in Fig. 3 of Ref.~\cite{BES12}, the $N{\bar N}$ amplitude
in the $I=1$ channel, which successfully describes the rather 
strong enhancement detected in the reaction $J/\psi {\to}\gamma p{\bar p}$ 
\cite{Sibirtsev05}, overestimates the energy dependence seen in the
$\psi(3686)$ case, cf. the dashed curve. 
On the other hand, the result based on the isospin averaged amplitude,
$(T_{I=1}+T_{I=0})/2\equiv T_{p \bar p}$, 
shown in Fig.~\ref{fig1} by the solid line, 
agrees rather nicely with the energy dependence found in the experiment.
With an appropriately chosen normalization, cf. Eq.~(\ref{fsi}), 
the data are well reproduced from the $p{\bar p}$ threshold up to excess 
energies of about 150 MeV. In particular, the $\chi^2$ is 8.7 for
the 15 data points shown in Fig.~\ref{fig1} while it is 22.2, i.e.
more than twice as large, for the pure $I=1$ amplitude. 
We also include the result based on FSI effects due to 
the $p{\bar p}$ amplitude in the $I=0$ channel alone (dash-dotted curve) and we
indicate the pure phase-space behaviour by the dotted curve. The latter
is obtained by using a constant amplitude $A$ in Eq.~(\ref{trans}). 
The $\chi^2$ value for the pure $I=0$ amplitude is 11.9. The one
for the phase-space curve amounts to 60 which is a clear indication that
the measured invariant mass spectrum does not exhibit a phase-space 
behaviour near threshold. 
All those curves are normalized to the solid curve at $M_{p\bar p}-2m_p\approx$ 60 MeV
in order to facilitate a comparison of the differences in the energy
dependence. 

Based on those findings we do not see any need here to invoke further more
substantial contributions coming from any $f_2(1910)$ or $f_2(1950)$ mesons, say, as 
done in Refs.~\cite{BES12,CLEO10}, in order to explain the data. 

In Fig.~\ref{fig2} our results are compared with the data obtained by
the CLEO Collaboration \cite{CLEO10}. Given the poorer statistics it is 
difficult to infer the actual energy dependence of the $p{\bar p}$ invariant 
mass. Still, the solid curve corresponding to the FSI effects 
based on $T_{p \bar p} = (T_{I=1}+T_{I=0})/2$, again appropriately 
normalized, comes closest to the trend exhibited by the 
data. But in terms of the $\chi^2$ the differences are marginal. All
results including FSI effects yield $\chi^2$/data $\approx$ $20/30$ while
the pure phase space amounts to $30/30$.

\begin{figure}[t]
\psfig{file=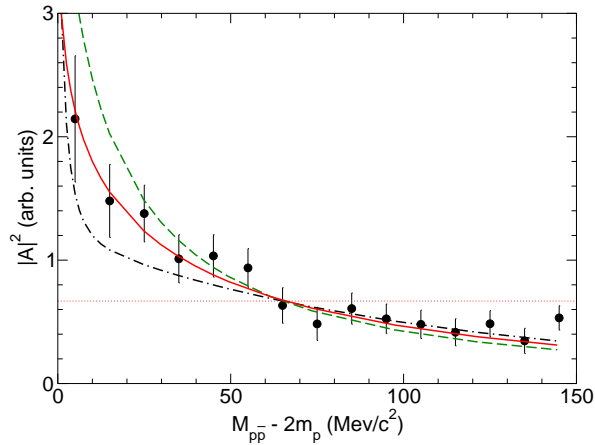,height=8.5cm,angle=-90}
\caption{
Invariant $\psi(3686) {\to}\gamma p{\bar p}$ amplitude $|A|^2$ as a
function of the $p{\bar p}$ mass. The circles symbolize the experimental
values of $|A|^2$ extracted from the BES data \cite{BES12}
via Eq.~(\ref{trans}). The curves are the appropriately normalized
scattering amplitude squared, $|T|^2$, predicted by the $N\bar N$
model A(OBE) \cite{Hippchen,Mull} for the $^{1}S_0$ partial wave.
For notation of curves, see Fig.~\ref{fig1}.
}
\label{fig3}
\vspace*{-5mm}
\end{figure}

Finally, in Fig.~\ref{fig3} the results for BESIII 
are displayed again, however this time in
terms of the modulus squared of the amplitude $A$. Here the curves 
correspond directly to the (appropriately normalized) scattering amplitude 
squared ($|T|^2$) predicted by the $N\bar N$ model A(OBE) \cite{Hippchen,Mull} 
for the $^{1}S_0$ partial wave. The symbols indicate the experimental
values of $|A|^2$, obtained from the BESIII data \cite{BES12} via dividing
the latter by the kinematical factors according to Eq.~(\ref{trans}). 

In summary, we have analyzed 
the near-threshold behavior of the $p{\bar p}$ invariant mass spectrum
from the $\psi(3686){\to}\gamma p{\bar p}$ decay reported recently 
by the BESIII Collaboration within the Watson-Migdal approach. 
Although in this reaction there is definitely
an enhancement in the near-threshold region as compared to the phase-space
behavior,  it is much less pronounced than what was found for the 
corresponding reaction $J/\psi{\to}\gamma p{\bar p}$. 
The enhancement is nicely reproduced by the $p{\bar p}$ final-state interaction 
based on the isospin averaged $^{1}S_0$ partial-wave amplitude as given by the 
J\"ulich $N\bar N$ model. In particular, any more substantial contributions from 
tensor mesons like $f_2(1910)$ or $f_2(1950)$, as 
advocated in earlier works \cite{CLEO10,BES12}, are not required.
 
Note that we have used here the same $N\bar N$ amplitudes as in our study of the
$J/\psi$ decay \cite{Sibirtsev05}. In the $J/\psi$ case the FSI provided by the 
$I=1$ component alone led to an agreement with the measured near-threshold $p{\bar p}$ 
invariant mass spectrum. Clearly, the mechanisms for the decay of the $J/\psi$ and 
$\psi(3686)$ mesons into $\gamma p{\bar p}$ should be different so that different
admixtures of the two isospin components in the final $p{\bar p}$ state have to be
expected. Only dedicated microscopic calculations, which hopefully will be performed
in the future, can allow to shed light on the details of the reaction mechanisms. 

\acknowledgments{
This work is supported in part by the DFG and the NSFC through
funds provided to the Sino-German CRC~110 ``Symmetries and
the Emergence of Structure in QCD'', and by the European
Community-Research Infrastructure Integrating Activity ``Study of
Strongly Interacting Matter'' (acronym HadronPhysics3).
}

\end{document}